\documentclass[letterpaper, 10pt, conference]{ieeeconf}      
\usepackage{amsmath, amssymb}
\usepackage{graphicx}
\usepackage{subfigure}
\usepackage[noadjust]{cite}
\usepackage{color}
\usepackage{amsmath, amssymb}
\usepackage{graphicx}
\usepackage{subfigure}
\usepackage[noadjust]{cite}
\usepackage{color}
\usepackage[noadjust]{cite}
\usepackage{color}
\usepackage{wrapfig}

\usepackage{algorithmic}
\usepackage{algorithm}

\usepackage{hyperref}
\usepackage{mathrsfs}
\usepackage{cite}
\IEEEoverridecommandlockouts                              
\title{\LARGE \bf
Barrier Function-based Safe Reinforcement Learning for Emergency Control of Power Systems
}

\author{Thanh Long Vu,~\IEEEmembership{Member,~IEEE,} Sayak Mukherjee,~\IEEEmembership{Member,~IEEE,} \\ Renke Huang,~\IEEEmembership{Senior Member,~IEEE,} 
and~Qiuhua Huang,~\IEEEmembership{Member,~IEEE}
\thanks{T. L. Vu, S. Mukherjee, R. Huang, Q. Huang are with the Pacific Northwest National Laboratory, 902 Battelle Blvd, Richland, WA 99354 USA. Corresponding author: Thanh Long Vu. E-mail: \texttt{thanhlong.vu@pnnl.gov}. 
}}

\begin{document}

\maketitle
\thispagestyle{empty}
\pagestyle{empty}

\maketitle
\begin{abstract}
    Under voltage load shedding has been considered as a standard and effective measure to recover the voltage stability of the electric power grid under emergency and severe conditions. However, this scheme usually trips a massive amount of load which can be unnecessary and harmful to customers. Recently, deep reinforcement learning (RL) has been regarded and adopted as a promising approach that can significantly reduce the amount of load shedding. However, like most existing machine learning (ML)-based control techniques, RL control usually cannot guarantee the safety of the systems under control. In this paper, we introduce a novel safe RL method for emergency load shedding of power systems, that can enhance the safe voltage recovery of the electric power grid after experiencing faults. Unlike the standard RL method, the safe RL method has a reward function consisting of a Barrier function that goes to minus infinity when the system state goes to the safety bounds. Consequently, the optimal control policy, that maximizes the reward function, can render the power system to avoid the safety bounds. This method is general and can be applied to other safety-critical control problems. Numerical simulations on the 39-bus IEEE benchmark is performed to demonstrate the effectiveness of the proposed safe RL emergency control, as well as its adaptive capability to faults not seen in the training. 
\end{abstract}

\section{Introduction}

Significant investment and efforts have been devoted to hardening grid infrastructures in the U.S. Preventive control measures, such as out-of-merit generation dispatch, have been widely adopted to ensure adequate security margins. However, on any given day, about 500,000 customers are without power for 2 hours or more. Several large blackouts have occurred in the U.S. in the last 20 years. As such, emergency control, i.e., quick actions to recover the stability of a power grid  under critical contingency, is more frequently required.
Currently, emergency control of power grids is largely based on remedial actions, special protection schemes (SPS), and load shedding \cite{Vittal2003}, which aim to quickly rebalance power and hopefully stabilize the system. These emergency control measures historically make the electrical power grid reasonably stable to disturbances. However, with the increased uncertainties and rapidly changing operational conditions in power systems, the existing emergency control methods have exposed outstanding issues in terms of either speed, adaptiveness, or scalability. 

For the emergency voltage control problem, load shedding is well known, among the measures of emergency voltage control, as a standard and effective countermeasure \cite{Taylor92}. It has been widely adopted in the industry, mostly in the form of rule-based undervoltage load shedding (UVLS). The UVLS relays are usually employed to shed load demands at substations in a step-wise manner if the monitored bus voltages fall below the predefined voltage thresholds. ULVS relays have a fast response, but do not have communication or coordination among substations that can lead to unnecessary load shedding  at affected substations \cite{Bai11}. 
In parallel, to reduce the dependence on the model, multiple data-driven approaches have been investigated to tackle voltage control problems.
In \cite{Genc2010_DT},  a decision tree based approach was introduced for preventive and corrective control actions. In \cite{Qiao2020hierarchical_LS}  a hierarchical extreme-learning machine based algorithm was presented for load shedding against fault-induced delayed
voltage recovery (FIDVR) events. 

Recently, reinforcement learning (RL) based approach has been successfully developed to solve the emergency voltage control problem \cite{huang2019loadshedding_DRL, zhang2018loadshedding_DRL, huang2020accelerated}. In the RL approach, the machine agent interacts with the system/environment, observes the resulting system state and the corresponding reward (which is suitably defined to encourage the voltage stability constraint), and updates the control actions in a way to maximize the reward \cite{huang2019loadshedding_DRL}. In comparison to the rule-based UVLS, the RL-based load shedding can significantly reduce the amount of loads that need to be shed in order to recover the voltage stability of the system \cite{huang2020accelerated}.
Dominantly considered in the existing literature, Markov decision process (MDP) based framework was utilized to develop a varieties of RL algorithms, many of those were investigated for solving  emergency voltage control problem. In \cite{huang2019loadshedding_DRL}, we have designed a deep Q learning based RL control  for emergency load shedding in response to voltage stability issues of the electric power grid. Subsequently, an accelerated Augmented Random Search (ARS) algorithm was introduced in \cite{huang2020accelerated} to quickly train the neural network-based control policy in a parallel computing mechanism. This ARS method, sketched in Section \ref{sec.ARS}, is implemented on a high-performance computing platform in a novel nested parallel architecture using the Ray framework. The  nested parallel architecture allowed for parallelizing power grid dynamic simulations in the ARS training and  helped exploring the parameter space of the control policy efficiently and adapting ARS agent to multiple tasks.

Though multiple reinforcement learning methods were developed in the literature, the safety of the system under reinforcement learning control is paid relatively less attention.  Recently, we introduced an optimization constrained-based safe RL method \cite{vu2020safe}, in which the machine learning agent will search for the optimal control policy that maximizes the reward function, while obeying the safety constraint. In this paper, we develop a safe ARS algorithm that can enhance the safety of the power grid during training, while inheriting all the advantages of the standard ARS algorithm. In this safe ARS method, as described in Section \ref{sec.safeARS}, the reward function is added a Barrier function that will go to minus infinity when the system state tends to the safety bounds. As the RL agent learns the optimal control policy that maximizes the reward function, it searches over the control policy space and selects the beat-performance policies in each iteration where the reward function is highest. Also, the control policy is updated by the gradient method applied on the reward function. As such, the reward function is increasing in expectation during the update of control policy. Therefore, in the searching of optimal control policy, the reward function and the Barrier function cannot tend to minus infinity. Hence, the safety bounds of system state are not violated. 

The difference of this paper in comparison with \cite{vu2020safe} is that \cite{vu2020safe} solves a constrained optimization problem to find the optimal control policy, while this paper does not solve a constrained optimization problem. Instead,  a Barrier function is included into the reward function to guide the searching of the optimal control policy. In this sense, the reward function serves as the control Lyapunov function in control theory. In a related work \cite{cheng2019endtoend}, model-based control Barrier Lyapunov function was used to design a control to compensate for the model-free reinforcement learning control in order to ensure the safety of the system. One limitation of this work lies on the limited scalability of the on-line learning of unknown system dynamics, and thus, this method was only demonstrated in simple lower-order systems. The safe ARS method we present in this paper does not involve any model-based design, but only searches over the control policy space. As such, the proposed method is applicable to large scale systems. Another similar work was presented in \cite{choi2020reinforcement} where  a unified RL-based framework was used to learn the dynamic uncertainty in the control Lyapunov function, control Barrier function, and other dynamic control affine constraints altogether in a single learning process. Again, the distinction of our work is that the Barrier function was included in the reward function to guide the control learning process, without any model knowledge.

Section II introduces the RL-based grid emergency voltage control problem. The Barrier function based safe reinforcement learning methodology is described in Section III. We demonstrate the effectiveness of the proposed safe ARS-based emergency load shedding on the 39-bus IEEE testcase in Section \ref{sec.results}. Furthermore, we show that in comparison to the standard ARS-based load shedding \cite{huang2020accelerated}, the Barrier function-based safe ARS load shedding can be more adaptive to contingencies not seen in the training, indicating its promise in ensuring the safe operation of actual power systems in deployment.

\section{Reinforcement Learning-based Emergency Load Shedding}
\label{sec.ARS}

In the general RL framework, a RL agent interacts with the system/ environment, observes the resulting system state and the corresponding reward, and updates the control actions in a way to maximize the expected reward (or the expected sum of discounted rewards). Mathematically, we define a policy search problem in a (partially observable) MDP defined by a tuple $(S,A, \mathcal{P},r,\gamma)$ \cite{SuttonBarto}. The state space $S \subseteq \mathbb{R}^n$ and action space $ A \subseteq \mathbb{R}^m$ could be continuous or discrete. In this paper, both of them are continuous. The environment transition function $\mathcal{P}: S \times A \to S$  is the probability density of the next state $s_{t+1} \in S$ given the current state $s_t \in S$ and action $a_t \in A$. At each interaction step, the environment returns a real reward $r: S \times A \to R$. $\gamma \in (0,1)$ is the discount factor. 
The transition probability function $\mathcal{P}: S \times A \to S$ for the power grid problem characterizes the stochastic transition of the grid states during the dynamic events. This framework can also encompass any perturbations caused due to load variations or renewable fluctuations. 



Applying reinforcement learning into the emergency voltage control of power systems, we consider the following safety bounds, observation space and action space.

\begin{figure}[t!]
    \centering
    \includegraphics[width = 1.0\linewidth, height = 5.5 cm]{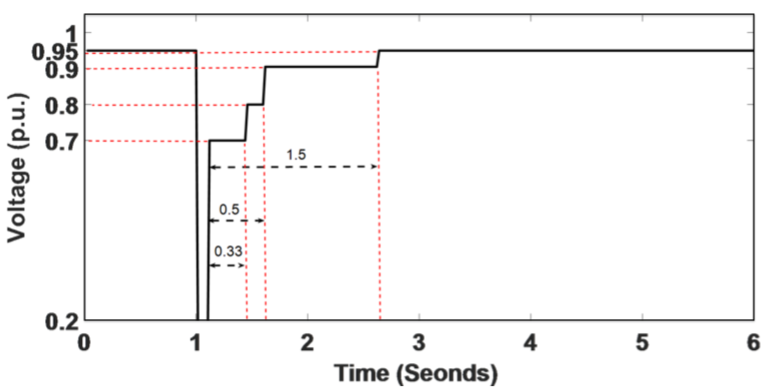}
    \caption{Safety bounds of voltages due to the transient voltage recovery criterion for transmission system}
    \label{fig.requirement}
    \vspace{-.5 cm}
    \end{figure}

{\bf Safety bounds:} In the load shedding problem, the objective is to recover the voltages of the electric power grid after the faults so that the post-fault voltages  will recover according to the standard [14] showed in Figure \ref{fig.requirement}. 
In particular, the standard requires that, after fault clearance, voltages should return to at least $0.8, 0.9$, and $0.95$ p.u. within $0.33s, 0.5s$, and $1.5s$. The states of the power grid should obey these time-dependent bounds. Let us denote the safety set $\mathcal{C}_i \subset \mathbb{R}^n$ for the $i^{th}$ time interval $t \in T_i$ after faults, then we would require,
\begin{align}
    s_t \in \mathcal{C}_i, \;\; t \in T_i.
\end{align}

\textbf{Observation space:} Accessing all the dynamic states of the power system is a difficult task, and the operators can only measure a limited number of states and  outputs. For the voltage control problem, the bus voltage magnitudes $V(t)$ 
are easily measurable, and therefore, considered in the the observation space. Please note that with slight abuse of notations, we continue to denote partially observable states or the outputs by $s_t$.

    \textbf{Action space:} We consider controllable loads as actuators where load shedding locations are generally set by the utilities by solving a rule-based optimization problem for secure grid operation. We consider the operator can shed upto $20\%$ of the total load at a particular bus at any given time instant. The action space is continuous with $[-0.2,\;0]$ range where $-0.2$ denotes the $20\%$ load shedding.

{\bf Safe RL control problem:} The control objective is to design reinforcement learning algorithm in which the RL agent will learn, over the action space, an optimal control policy that maximizes the expected reward function, while obeying the safety bounds of power system voltages.  

In our previous work, we have designed a deep Q learning based RL control \cite{huang2019loadshedding_DRL} for emergency load shedding to address this problem. Subsequently, we developed an accelerated Augmented Random Search (ARS) algorithm in \cite{huang2020accelerated} to quickly train the neural network-based control policy in a parallel computing mechanism. In the ARS algorithm in \cite{huang2020accelerated}, the ARS agent performs parameter-space exploration and estimates the gradient of the expected reward using sampled rollouts to update the control policy. In particular, for the above load shedding problem, the ARS agent's objective is to maximize the expected reward, where
the reward $r_t$ at time $t$ was defined as follows:
\begin{align}
\label{eq.reward}
r(t) = 
\begin{cases}
&-1000  {\;\;\; \emph{if}\;\;\;} V_i(t)<0.95,\;\;\; T_{pf} +4<t \\
& c_1 \sum_i \Delta V_i(t) - c_2 \sum_j \Delta P_j (p.u.) - c_3 u_{ivld}, \\&{\;\;\; \emph{\mbox{otherwise}},}
\end{cases}
\end{align}
where,
\vspace{-.5 cm}
\footnotesize
\begin{align}
\Delta V_i(t) =
\begin{cases}
&\min \{V_i(t) - 0.7, 0\},  \emph{\;if\;} t \in (T_{pf}, T_{pf} +0.33) \\
&\min \{V_i(t) - 0.8, 0\}, \emph{\;if\;} t \in (T_{pf} +0.33, T_{pf} +0.5 ) \\
&\min \{V_i(t) - 0.9, 0\} , \emph{\;if\;}  t \in (T_{pf} +0.5, T_{pf} +1.5) \\
&\min \{V_i(t) - 0.95, 0\} , \emph{\;if\;} t \in t > T_{pf} +1.5.
\end{cases}\nonumber
\end{align}
\normalsize

In the reward function \eqref{eq.reward}, $T_{pf}$ is the time instant of fault clearance;
$V_i(t)$ is the voltage magnitude for bus $i$ in the power
grid;  $\Delta P_j (t)$ is the
load shedding amount in p.u. at time step $t$ for load bus $j$; 
invalid action penalty $u_{ivld}$ if the DRL agent still provides
load shedding action when the load at a specific bus has already been shed to zero at the previous time step when
the system is within normal operation. $c_1 , c_2$, and $c_3$ are the weight factors for the above three parts.

 Furthermore, to scale up the ARS algorithm and reduce the training time, we accelerated it by leveraging its inherent parallelism and implementing it on a high-performance computing platform in a novel nested parallel architecture using the \textit{Ray} framework. This architecture allowed parallelizing power grid dynamic simulations in the ARS training and helped exploring the parameter space of the control policy efficiently and adapting ARS to multiple tasks \cite{huang2020accelerated}.

\section{Barrier Function-based Safe ARS for Emergency Load Shedding}
\label{sec.safeARS}

\subsection{Problem reformulation with Barrier function} 

In the proposed method, the reward function is included with a Barrier function that will go to minus infinity when the system state tends to the safety bounds.
Accordingly, for the voltage safety requirement in Figure \ref{fig.requirement}, the following time-dependent Barrier function can be used:

\begin{align}
\label{safeVol}
   B(s_t,t)= \begin{cases}
   & \sum_i 1/(V_i(t)-0.7)^2) \\& {\;\;\; \emph{if}\;\;\;} T_{pf}<t<T_{pf}+0.33, \\
   & \sum_i 1/(V_i(t)-0.8)^2) \\& {\;\;\; \emph{if}\;\;\;} T_{pf}+0.33<t<T_{pf}+0.5, \\
   & \sum_i 1/(V_i(t)-0.9)^2)  \\&{\;\;\; \emph{if}\;\;\;} T_{pf}+0.5<t<T_{pf}+1.5, \\
   & \sum_i 1/(V_i(t)-0.95)^2)  \\&{\;\;\; \emph{if}\;\;\;} t>T_{pf}+1.5. 
   \end{cases}
\end{align}
where $V_i(t)$ is the bus voltage magnitude for bus $i$ in the power
grid. The Barrier function $B(s_t,t)$ will go to infinity when the voltages tends to the safety bounds in Figure \ref{fig.requirement}.

Now, the reward function we consider in the safe ARS algorithm is as follows:
\begin{align}
\label{reward}
    R(t) = r(t) - c_4 B(s_t,t),
\end{align}
where $r(t)$ is defined in \eqref{eq.reward}, $B(s_t,t)$ is defined in \eqref{safeVol}, and $c_4>0$ is a weight factor. \\
\textbf{Reformulated safe RL problem:} 
We consider the following optimization problem:
\begin{align}
    & \mbox{maximize}_{\theta}\;\; \mathbb{E}(\sum_t R(t)), \nonumber \\
     & \mbox{s.t.} \;\; s_{t+1} \sim P(s_{t+1}\;| s_{t}, a_t, d_t), \label{grid_dynamics} \nonumber \\
     & a_t = \pi_{\theta}(s_t), \;\; a_t \in [a_t^{\mbox{min}}, a_t^{\mbox{max}}],
\end{align}
where $\pi_{\theta}$ is the non-linear control policy to be designed via the safe RL algorithm. In our learning design, we model the control policy as a long short term memory (LSTM) network \cite{Schmidhuber1997_LSTM}  due to its capability of automatically learning to capture the temporal dependence over multiple time steps. The actions generated via the non-linear control policy are also considered to be constrained within practically allowable range. The variable $d_t$ denotes the disturbance input to the power grid. Please note that the actual grid dynamics is much more complex than the notation simplicity we used for \eqref{grid_dynamics}, as the grid characteristics can be described by a set of dynamic state variables (such as generator angles, frequencies etc.), and a set of algebraic variables (bus voltages for example) \cite{sauer1998power}. For the RL-based control design, we use a subset of those states and algebraic variables as our observables. In our numerical grid simulator, we have performed the detailed dynamic simulations on a benchmark power grid model.

\subsection{Barrier function-based safe ARS algorithm}

\begin{algorithm}[]
\caption{Barrier function-based safe ARS:}
\begin{algorithmic}
\STATE 1. \textbf{Hyperparameters:} Step size $\alpha$, number of policy perturbation directions per iteration $N$, standard deviation of the exploration noise $\nu$, number of top-performing perturbed directions selected for updating weights $b$, number of rollouts per perturbation direction $m$. Decay rate $\epsilon$. 
\STATE 2. \textbf{Initialize:} Policy weights $\theta_0$ with small random numbers; initialize the running mean of observation states $\mu_0 = 0 \in R^n$ and the running standard deviation of observation states $\Sigma_0 = I_n \in R^n$, where $n$ is the dimension of observation states, the total iteration number $H$.
\FOR{ iteration $t \leq H$}
\STATE 3. Sample $N$ number of random directions $\delta_{1},\dots,\delta_{N} \in \mathbb{R}^{n_{\theta}}$ with the same
dimension as policy weights $\theta$.
\FOR{each $\delta_{i}, i=1,\dots, N$}
\STATE 4. Add perturbations to policy weights: $\theta_{ti+} = \theta_{t - 1} +
\nu \delta_i$ and $\theta_{ti-} = \theta_{t - 1} - \nu \delta_i$
\STATE 5. Do total $2m$ rollouts (episodes) denoted by $R_{p\in T} (.)$
for different tasks $p$ sampled from task set $T$ corresponding to $m$ different faults with the $\pm$ perturbed policy weights. Calculate the average rewards of $m$ rollouts as the rewards for $\pm$ perturbations, i.e., $\bar{R}_{ti +}$ and $\bar{R}_{ti -}$ are 
\begin{align}
    &\bar{R}_{ti+} = \frac{1}{m} R_{p\in T}(\theta_{ti+},\mu_{t - 1},\Sigma_{t - 1}),\\
    &\bar{R}_{ti-} = \frac{1}{m} R_{p\in T}(\theta_{ti-},\mu_{t - 1},\Sigma_{t - 1})
\end{align}
where the reward function $R$ is defined as the combined reward function in \eqref{reward}.

\STATE 6. During each rollout, states $s_{t,k}$ at time step $k$ are first normalized and then used as the input for inference with policy $\pi_{\theta_t}$ to obtain the action $a_{t,k}$, which is applied to the environment and new states $s_{t,k+1}$ is returned, as shown in (3). The running mean $\mu_t$ and standard deviation $\Sigma_t$ are updated with $s_{t,k+1}$
\begin{align}
    &s_{t,k} = \frac{s_{t,k} - \mu_{t-1}}{\Sigma_{t-1}},\\
    & a_{t,k} = \pi_{\theta_t}(s_{t,k}),\\
    & s_{t,k+1} \gets \mathcal{P}(s_{t,k},a_{t,k}) ,
\end{align}
\ENDFOR
\STATE 7. Sort the directions based on $\max ( \bar{R}_{ti + } , \bar{R}_{ti - } )$, select
top $b$ directions, calculate their standard deviation $\sigma_b$
\STATE 8. Update the policy weight:
\begin{align}
    \theta_{t+1} = \theta_{t} + \frac{\alpha}{b  \sigma_{b}} \sum_{i=1}^{b} (\bar{R}_{ti+} - \bar{R}_{ti-})\delta_{i}
\end{align}
\STATE 9. Step size $\alpha$ and standard deviation of the exploration noise $\nu$ decay with rate $\epsilon$: $\alpha \gets \epsilon \alpha, \nu\gets \epsilon \nu$.
\ENDFOR
\RETURN 10.  Return $\theta$.
\end{algorithmic}
\end{algorithm}

\subsubsection{Algorithmic overview}
Algorithm $1$ presents the steps to compute the optimal control policy by the Barrier function-based safe ARS method. This algorithm has the following salient characteristics. 

\begin{itemize}
    \item Similar to the standard ARS algorithm \cite{huang2020accelerated}, the safe ARS learner is an actor at the top to delegate tasks and collect returned information, and controls the update of policy weights $\theta$.
    \item The learner communicates with subordinate workers and each of these workers is responsible for one or more perturbations (random search) of the policy weights as in Step 4.
    \item In Steps 7, the ARS learner combines the results from each worker calculated in Step 5 (which include the average reward of multiple rollouts), sorts the directions according to the reward, selects the  best-performing directions.
    \item Then, in Step 8, the ARS learner updates the policy weights centrally based on the perturbation results from the top performing workers.
    \item The workers do not execute environment rollout tasks by themselves. They spawn a number of actors and assign these tasks to these subordinate  actors. Note that each worker needs to collect the rollout results from multiple tasks inferring with the same perturbed policy, and each actor is only responsible for one environment rollout with the specified task and perturbed policy sent by its up-level worker. For the environment rollouts, power system dynamic simulations are performed in parallel. 
\end{itemize}

We note that in comparison with the constrained optimization-based safe ARS algorithm in \cite{vu2020safe}, Algorithm 1 is simpler in the sense that we do not need to check the safety of the system in each iteration in order to update the safety multiplier in the reward function.

\subsubsection{Safety Considerations}
In this safe ARS algorithm, as the ARS agent learns the optimal control policy that maximizes the reward function, in each iteration it will search over the control policy space and select, among several directions, the best-performance policies where the reward function is highest. Also, when going to the next iteration, the control policy is updated by the surrogate gradient-like method on the reward function. As such, the reward function is increasing in expectation during the update of control policy. Hence, during the exploration and update of the control policy, the reward function cannot tend to minus infinity. As a result, the Barrier function cannot tend to minus infinity during the exploration and update of the control policy. Therefore, the safety bounds of system state are not violated during the searching process of the optimal control policy. 

Mathematically, in the ARS-based learning process for the optimum control policy, the expected reward is lower bounded. This means that, along the trajectory of the agent state, there is only a zero-measure set of samples in which the reward function can go to infinity. Hence, the reward function is bounded almost surely, i.e., $\mathbf{P}\{R_t >-\infty, \forall t\}=1$. Hence, the Barrier function is also bounded almost surely along the trajectory $s(t).$ 
We note that when $s(t)$ goes through the safety bounds, then the Barrier function will go to minus infinity. Therefore, we can conclude that the voltages will not violate the safety bounds almost surely.

\section{Test Results}
\label{sec.results}
\begin{figure}[t!]
    \centering
    \includegraphics[width = .8\linewidth, height = 5.5 cm]{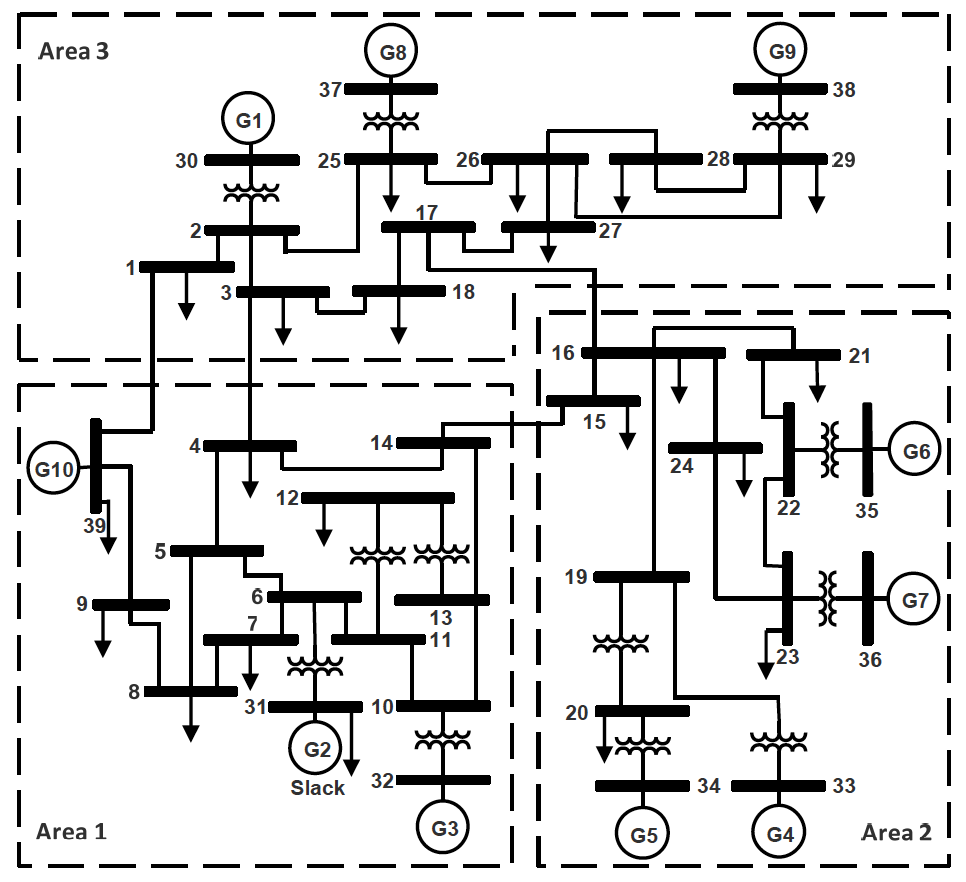}
    \caption{IEEE 39-bus system}
    \label{fig:39bus}
    \vspace{-.5 cm}
    \end{figure}

We perform simulations in the IEEE benchmark $39-$bus, $10-$generator model, as shown in Fig. \ref{fig:39bus}. The simulations were undertaken in a Linux mainframe with $27$ cores. The power system simulator runs using GridPack\footnote{https://www.gridpack.org}, 
and the safe deep RL algorithm is implemented in a separate platform using python. A software setup has been built such that the grid simulations in the GridPack and the RL iterations in the python can communicate.  

\begin{figure}[]
    \centering
    \includegraphics[width = 1\linewidth, height = 5.5 cm]{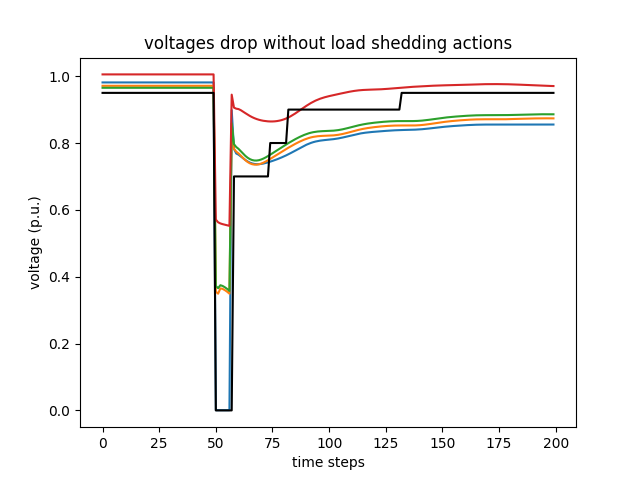}
    \caption{Voltage drops after fault happens at bus 4 without any emergency control measures}
    \label{fig:nocontrol}
    \vspace{-.5 cm}
\end{figure}
\begin{figure}[t]
    \centering
    \includegraphics[width = 1\linewidth, height = 5.5 cm]{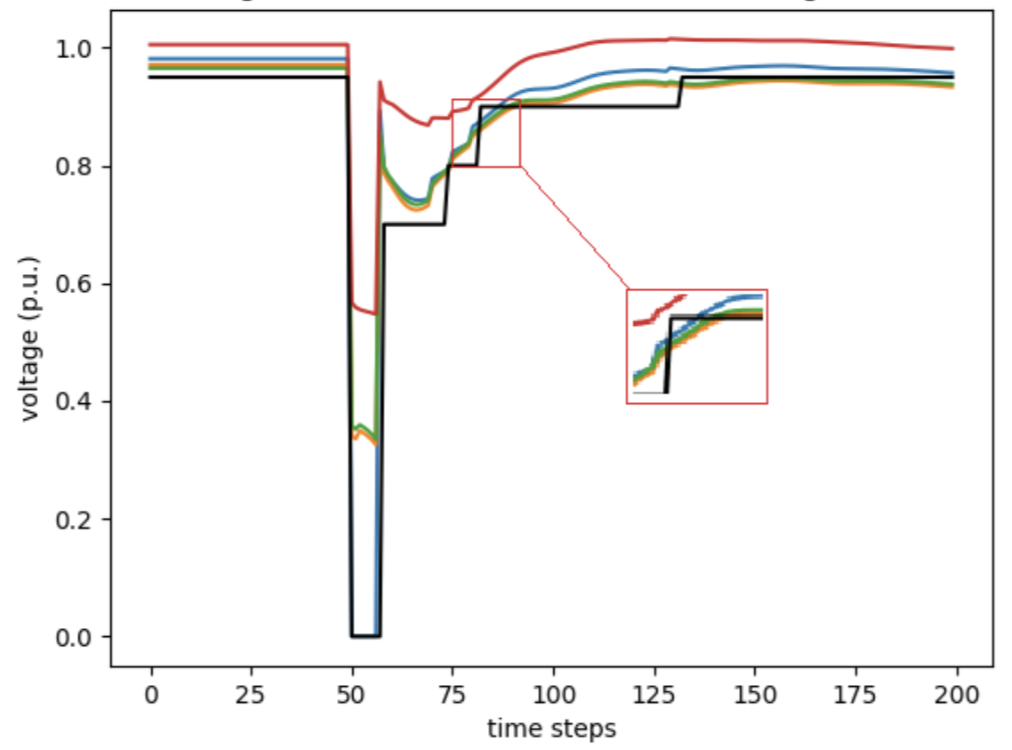}
    \caption{Voltage profile under the standard ARS-based emergency load shedding \cite{huang2020accelerated} response to the fault that happens at bus 4}
    \label{fig:ARS4}
    \vspace{-.5 cm}
\end{figure}
\begin{figure}[t]
    \centering
    \includegraphics[width = 1\linewidth, height = 5.5 cm]{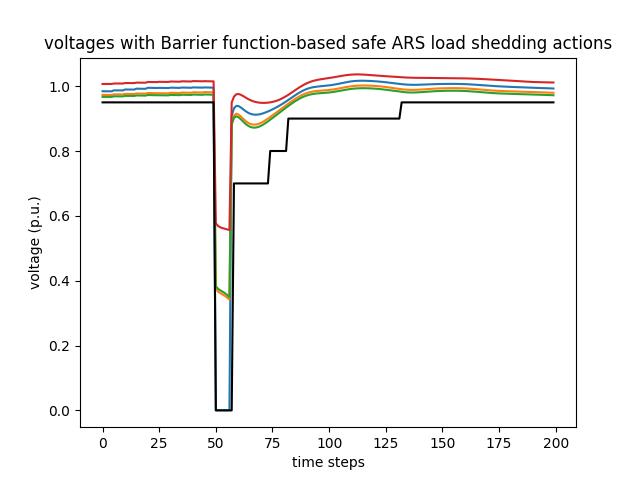}
    \caption{Voltage profile under the Barrier function-based safe ARS emergency load shedding response to the fault that happens at bus 4.}
    \label{fig:safeARS4}
    \vspace{-.5 cm}
\end{figure}

The performance of our Barrier function-based safe ARS algorithm has been tested on the above mentioned IEEE $39$-bus system to learn a closed-loop non-linear control policy. The policy generates required optimized load shedding actions at buses $4, 7,$ and $18$ to avoid the FIDVR event and meet the voltage recovery requirements shown in Fig. 1. 
Observations included voltage magnitudes at buses $4, 7, 8,$ and $18$ as well as the remaining fractions of loads served by buses $4, 7$ and $18$. The control action for buses $4, 7,$ and $18$ at each action time step was a continues variable from $0$ (no load shedding) to $-0.2$ (shedding $20\%$ of the initial total load at the bus). We fix a the task set T consisting of nine different tasks (fault scenarios) for the training purposes. The fault scenarios began with the flat start in the dynamic simulation. At $1.0$ s, we apply short circuit faults at one of the bus $4, 15$, or $21$ with a fault duration of $0.0$ s (no fault), $0.15$ s, or $0.28$ s and the fault was self-cleared. 
In \cite{huang2020accelerated}, it is shown that the standard ARS can offer effective load shedding to safeguard the grid in many contingencies.
In this paper, to test the effectiveness of the safe ARS-based load shedding in comparison with the standard ARS-based load shedding, relatively severe faults that can harm the system are being considered. As such, we consider relatively large fault duration (e.g., $0.28$ seconds fault caused severe disturbance in the grid). 


\begin{figure}[t]
    \centering
    \includegraphics[width = 1\linewidth, height = 5.5 cm]{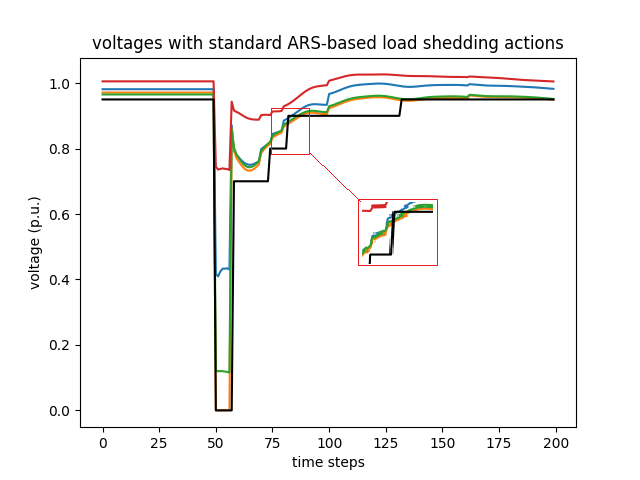}
    \caption{Voltage profile under the standard ARS-based emergency load shedding \cite{huang2020accelerated} response to the untrained fault that happens at bus 7}
    \label{fig:ARS}
    \vspace{-.5 cm}
\end{figure}

\begin{figure}[]
    \centering
    \includegraphics[width = 1\linewidth, height = 5.5 cm]{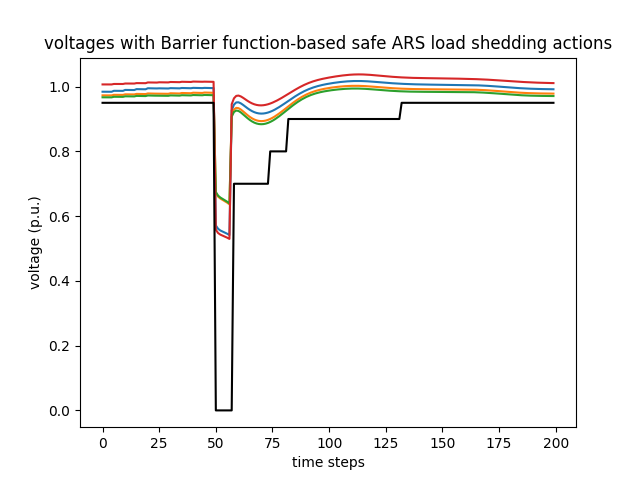}
    \caption{Voltage profile under the Barrier function-based safe ARS emergency load shedding response to the untrained fault that happens at bus 7. This shows that the Barrier function-based safe ARS algorithm can result in emergency control policy adaptive to the fault not seen during the training.}
    \label{fig:safeARS}
    \vspace{-.5 cm}
\end{figure}

To create a baseline, we consider a short-circuit fault that happens at bus $4$ at $1.0$s with a fault duration of $0.15$s, and then the fault is self-cleared. When there is no additional feedback control implemented in the system, we can observe voltage recovery profile largely violates the required recovery bounds as in Figure \ref{fig:nocontrol}.
For comparison with the standard ARS-based emergency control in \cite{huang2020accelerated}, we also train it with the same set of faults and deploy it to react the same fault at bus $4$. The comparison of performances for the standard ARS-based load shedding and safe ARS-based load shedding are depicted  in Figs. \ref{fig:ARS4} and \ref{fig:safeARS4}. 
The figure shows that the safe ARS-based load shedding can perform much better than the standard ARS-based load shedding in meeting the safety requirement described by the transient voltage recovery criterion depicted in Fig. 1. In particular, without any Barrier-function based RL design, the voltages at buses $7,8,$ and $18$ could not recover to $0.9$ p.u. within $0.5$s after the fault clearance, and the voltages at buses $8$ and $18$ do not recover to $0.95$ p.u. within $1.5$s after the fault-cleared time. Yet, the safe ARS-based load shedding makes the voltages at all buses $4,7,8,18$ recover well as required. 


To test the adaptation of the safe ARS, we consider a fault not encountered during the training, in which short-circuit fault happens at bus $7$ at $1.0$s with a fault duration of $0.15$s and then the fault self-cleared. The performances of standard ARS-based load shedding and safe ARS-based load shedding are depicted in Figs. \ref{fig:ARS} and \ref{fig:safeARS}. From these figures, we can observe that the safe ARS-based load shedding is better than the standard ARS-based load shedding not only  in meeting the safety requirement described by the transient voltage recovery criterion, but also in adapting to a fault not encountered during the training.

\section{Conclusions}

In this paper, we presented a highly scalable and safe deep reinforcement learning algorithm for power system voltage stability control using load shedding. This algorithm inherits the parallelism of the ARS algorithm and thereby achieve high scalability and applicability for power system stability control applications. Remarkably, by incorporating a Barrier function into the reward function, the safe ARS algorithm resulted in a control policy that could prevent the system state from violating the safety bounds, and hence, enhance the safety of the electric power grid during the load shedding. A small number of hyper-parameters makes this algorithm easy to tune to achieve good performance. Case studies on the IEEE $39$-bus demonstrated that safe ARS-based load shedding scheme successfully recovers the voltage stability of power systems even in events it did not see during the training. In addition, in comparison to the standard ARS-based load shedding, it showed advantages in both safety level and fault adaptability.

\bibliographystyle{IEEEtran}
\bibliography{ref.bib}
\end{document}